\magnification=1200
\output={\plainoutput}

\newcount\pagenumber
\newcount\questionnumber
\newcount\sectionnumber
\newcount\appendixnumber
\newcount\equationnumber
\newcount\referencenumber
\newcount\subsecnumber

\global\subsecnumber=1

\def\ifundefined#1{\expandafter\ifx\csname#1\endcsname\relax}
\def\docref#1{\ifundefined{#1} {\bf ?.?}\message{#1 not yet defined,}
\else \csname#1\endcsname \fi}


\newcount\linecount

\newcount\citecount
\newcount\localauthorcount 

\def\article{
\def\eqlabel##1{\edef##1{\sectionlabel.\the\equationnumber}}
\def\seclabel##1{\edef##1{\sectionlabel}}                  
\def\feqlabel##1{\ifnum\passcount=1
\immediate\write\crossrefsout{\relax}  
\immediate\write\crossrefsout{\def\string##1{\sectionlabel.
\the\equationnumber}}\else \fi }
\def\fseclabel##1{\ifnum\passcount=1
\immediate\write\crossrefsout{\relax}   
\immediate\write\crossrefsout{\def\string##1{\sectionlabel}}\else\fi}
\def\docite##1 auth ##2 title ##3 jour ##4 vol ##5 pages ##6 year ##7{
\par\noindent\item{\bf\the\referencenumber .}
 ##2, ##3, ##4, {\bf ##5}, ##6,     
(##7).\par\vskip-0.8\baselineskip\noindent{
\global\advance\referencenumber by1}}
\def\dobkcite##1 auth ##2 title ##3 publisher ##4 year ##5{
\par\noindent\item{\bf\the\referencenumber .}
 ##2, {\it ##3}, ##4, (##5).
\par\vskip-0.8\baselineskip\noindent{\global\advance\referencenumber by1}}
\def\doconfcite##1 auth ##2 title ##3 conftitle ##4 editor ##5 publisher ##6 
year ##7{
\par\noindent\item{\bf\the\referencenumber .}
##2, {\it ##3}, ##4,  {edited by: ##5}, ##6, (##7).
\par\vskip-0.8\baselineskip\noindent{\global\advance\referencenumber by1}}}

\def\normalarticlestyle{
\global\referencenumber=1
\def\eqlabel##1{\edef##1{\sectionlabel.\the\equationnumber}}
\def\seclabel##1{\edef##1{\sectionlabel}}                  
\def\feqlabel##1{\ifnum\passcount=1
\immediate\write\crossrefsout{\relax}  
\immediate\write\crossrefsout{\def\string##1{\sectionlabel.
\the\equationnumber}}\else \fi }
\def\fseclabel##1{\ifnum\passcount=1
\immediate\write\crossrefsout{\relax}   
\immediate\write\crossrefsout{\def\string##1{\sectionlabel}}\else\fi}
\def\z@{ 0pt}
\def\cite##1{\immediate\openin\bib=bib.tex\global\citecount=##1
\global\linecount=0{\loop\ifnum\linecount<\citecount \read\bib 
to\temp \global\advance\linecount by
1\repeat\immediate\write\reffs{\temp}\global\advance\referencenumber by1}
\immediate\closein\bib}
\def\docite##1 auth ##2 title ##3 jour ##4 vol ##5 pages ##6 year ##7{
\par\noindent\item{\bf\the\referencenumber .}
 ##2, ##3, ##4, {\bf ##5}, ##6,         
(##7).\par\vskip-0.8\baselineskip\noindent}
\def\dobkcite##1 auth ##2 title ##3 publisher ##4 year ##5{
\par\noindent\item{\bf\the\referencenumber .}
 ##2, {\it ##3}, ##4, (##5).
\par\vskip-0.8\baselineskip\noindent}
\def\doconfcite##1 auth ##2 title ##3 conftitle ##4 editor ##5 publisher ##6 
year ##7{
\par\noindent\item{\bf\the\referencenumber .}
##2, {\it ##3}, ##4,  {edited by: ##5}, ##6, (##7).
\par\vskip-0.8\baselineskip\noindent}}

\def\appendixlabel{\ifcase\appendixnumber\or A\or B\or C\or D\or E\or
F\or G\or H\or I\or J\or K\or L\or M\or N\or O\or P\or Q\or R\or S\or
T\or U\or V\or W\or X\or Y\or Z\fi}

\def\sectionlabel{\ifnum\appendixnumber>0 \appendixlabel
\else\the\sectionnumber\fi}

\def\beginsection #1
 {{\global\appendixnumber=0\global\advance\sectionnumber by1}\equationnumber=1
\par\vskip 0.8\baselineskip plus 0.8\baselineskip
 minus 0.8\baselineskip 
\noindent$\S$ {\bf \the\sectionnumber . #1}
\par\penalty 10000\vskip 0.6\baselineskip plus 0.8\baselineskip 
minus 0.6\baselineskip \noindent}

\def\subsec #1 {\bf\par\vskip8truept  minus 8truept
\noindent \ifnum\appendixnumber=0 $\S\S\;$\else\fi
$\bf\sectionlabel.\the\subsecnumber$ #1
\global\advance\subsecnumber by1
\rm\par\penalty 10000\vskip6truept  minus 6truept\noindent}

\def\beginappendix #1
{{\global\advance\appendixnumber by1}\equationnumber=1\par
\vskip 0.8\baselineskip plus 0.8\baselineskip
 minus 0.8\baselineskip 
\noindent
{\bf Appendix \appendixlabel . #1}
\par\vskip 0.8\baselineskip plus 0.8\baselineskip
 minus 0.8\baselineskip 
\noindent}

\def\no{\eqno({\rm\sectionlabel} .\the\equationnumber){\global\advance\equationnumber by1}}

\def\beginref #1 {\par\vskip 2.4 pt\noindent\item{\bf\the\referencenumber .}
\noindent #1\par\vskip 2.4 pt\noindent{\global\advance\referencenumber by1}}

\def\ref #1{{\bf [#1]}}


\normalarticlestyle
\advance\hsize by 2truept


\font\eightrm=cmr8
\font\sixrm=cmr6

\font\ninei=cmmi9
\font\eighti=cmmi8
\font\sixi=cmmi6
\skewchar\ninei='177 \skewchar\eighti='177 \skewchar\sixi='177

\font\ninesy=cmsy9
\font\eightsy=cmsy8
\font\sixsy=cmsy6
\skewchar\ninesy='60 \skewchar\eightsy='60 \skewchar\sixsy='60

\font\eightbf=cmbx8
\font\sixbf=cmbx6

\font\ninett=cmtt9
\font\eighttt=cmtt8

\hyphenchar\tentt=-1 
\hyphenchar\ninett=-1
\hyphenchar\eighttt=-1

\font\eightsl=cmsl8

\font\eightit=cmti8



\newskip\ttglue

\def\eightpoint{\def\rm{\fam0\eightrm}%
  \textfont0=\eightrm \scriptfont0=\sixrm \scriptscriptfont0=\fiverm
  \textfont1=\eighti \scriptfont1=\sixi \scriptscriptfont1=\fivei
  \textfont2=\eightsy \scriptfont2=\sixsy \scriptscriptfont2=\fivesy
  \textfont3=\tenex \scriptfont3=\tenex \scriptscriptfont3=\tenex
  \def\it{\fam\itfam\eightit}%
  \textfont\itfam=\eightit
  \def\sl{\fam\slfam\eightsl}%
  \textfont\slfam=\eightsl
  \def\bf{\fam\bffam\eightbf}%
  \textfont\bffam=\eightbf \scriptfont\bffam=\sixbf
   \scriptscriptfont\bffam=\fivebf
  \def\tt{\fam\ttfam\eighttt}%
  \textfont\ttfam=\eighttt
  \tt \ttglue=.5em plus.25em minus.15em
  \normalbaselineskip=9pt
  \let\sc=\sixrm
  \let\big=\eightbig
  \setbox\strutbox=\hbox{\vrule height7pt depth2pt width\z@}%
  \normalbaselines\rm}

\def\title{Modular invariance, lattice field theories and finite size corrections}

\headline={{\eightpoint \rm\ifnum\pageno=1\hfill\else\ifodd\pageno 
Nash and O' Connor \hfill \title 
\else \title \hfill Nash and O' Connor\fi\fi}}

\def\im{Im\,}

\def\mod{mod\,}
\def\longbar#1{\setbox1=\hbox{$#1$}
\setbox2=\vbox{\hrule width 0.8\wd1}
\raise0.5\ht1\hbox{${\lower\dp1\box2}\atop\box1$}}  
\def\mediumbar#1{\setbox1=\hbox{$#1$}
\setbox2=\vbox{\hrule width 0.6\wd1}
\raise0.5\ht1\hbox{${\lower\dp1\box2}\atop\box1$}}

\par\vfill
\centerline{\title}
\vskip1.25\baselineskip
\centerline{by}
\vskip1.25\baselineskip
\centerline{Charles Nash$^{*,{\dag}}$ and Denjoe O' Connor$^{\dag}$}
\par\vskip\baselineskip
\noindent
$^*$Department of Mathematical Physics\hfill 
$^{\dag}$School of Theoretical Physics
\par\noindent
St. Patrick's College\hfill              Dublin Institute for Advanced
Studies
\par\noindent
Maynooth\hfill                {\it and}\hfill  10 Burlington Road
\par\noindent
Ireland\hfill                            Dublin 4
\par\noindent
\null\hfill                              Ireland
\par\vskip3\baselineskip
\noindent
{{\bf Abstract}: We give a lattice theory treatment of certain one and two
dimensional quantum field theories. In one dimension we construct a combinatorial 
version of a non-trivial field theory on the circle which is of some independent interest
in itself while in two dimensions
we consider a field theory on a toroidal triangular lattice. 
\par
We take a continuous spin Gaussian model
on a toroidal triangular lattice with periods
$L_0$ and $L_1$  where the spins
carry a representation of the fundamental group of the torus labeled
by phases $u_0$ and $u_1$. We compute 
the {\it exact finite size and lattice corrections},
to the partition function $Z$, for arbitrary mass $m$ and phases $u_i$.
Summing $Z^{-1/2}$ over a specified set of phases gives the corresponding result for the Ising
model on a torus.
An interesting property of the model is that the limits $m\rightarrow0$ and $u_i\rightarrow0$  do not commute.
Also when $m=0$ the model exhibits a {\it vortex critical phase}
when at least one of the $u_i$ is non-zero.
In the continuum or scaling limit, for arbitrary $m$, the finite size
corrections to
$-\ln Z$ are {\it modular invariant} and for the critical phase are given by
elliptic theta functions. In the cylinder limit $L_1\rightarrow\infty$
the ``cylinder charge'' $c(u_0,m^2L_0^2)$ is a non-monotonic function of
$m$ that ranges from $2(1+6u_0(u_0-1))$ for $m=0$ to zero for
$m\rightarrow\infty$ but from which one can determine the central
charge $c$.
The study of the continuum limit of
these field theories provides a  kind of quantum theoretic analog of the  
link between certain  combinatorial and analytic topological quantities} 
\par\vskip\baselineskip\noindent
{\bf PACS numbers:} 05.40.+j, 05.50.+q, 64.60.-i, 11.15.Ha, 05.70.Fh, 75.10.Hk
\vfill\eject

\beginsection{Introduction}
In this paper we examine some discretely formulated quantum field
theories as well as successfully computing their continuum limits. We shall consider field 
theories
in one  dimension, over the circle $S^1$, and in two dimensions, over the torus $T^2$.  
However a crucial feature of the fields will be that they will {\it not} be singly-periodic and 
doubly-periodic {\it functions} over $S^1$ 
and $T^2$ respectively; instead they will be
{\it sections} of  appropriate bundles over  $S^1$ and $T^2$. These sections will then not 
be periodic but will possess a non-trivial {\it holonomy} when rotated round a non-trivial loop on
their base space. A brief summary of some of these results has been presented in \ref{1}.
\par
In our two dimensional model we shall be able to carry out an extensive calculation of finite size
effects which are associated with the onset of continuous phase transition.  
These effects provide corrections to the bulk system behaviour and play a vital r\^ole 
in our theoretical and experimental knowledge of  systems at and near criticality cf.
\ref{2}. Such effects have a complex interplay with the other features of these systems 
giving rise to phenomena such as crossovers from one characteristic behaviour to
another \ref{3}. Two dimensional physics has provided a framework where substantial progress
has been made in understanding such problems and considerable interest has been focused on them.
Some of these calculations use generalised Gaussian models \ref{4,5} and our work can be considered in a
similar spirit. 
\par
In one dimension we use a zeta function to regularise the theory and
also construct the theory discretely on a polygon. We  find that when the Van Vleck Morette 
determinant is included in the measure of the
discretely calculated partition function the massless case is independent of the number
$N$ of points in the polygon. In addition  
we find that the combinatorial and
continuum partition functions agree precisely and are equal to a power
of the Ray-Singer torsion for the appropriate flat bundle of sections
over $S^1$. 
\par
For our two dimensional theory we obtain the exact finite size corrections to the
free energy of our lattice model. These corrections
form a {\it modular invariant} expression and for the critical phase this expression
is expressible, in the continuum limit, 
in terms of elliptic theta functions and  the associated partition
function {\it holomorphically factorises} in the modular parameter $\tau$.
The model possesses two holonomy phases (cf. section 3) $u_0$ and $u_1$ 
and  a mass $m$; there is a critical vortex phase when the mass is zero and one of the phases is
non-zero. Modular invariance persists whether $m$ is zero or not. 
\par
The principal new results contained in this work are
\item{(i)}A demonstration that the combinatorial torsion arises naturally when the 
Van Vleck-Morette determinant is incorporated in the integration measure.
\item{(ii)}The determination of the exact partition function for a massive Gaussian 
model where the spin variables are sections over a generic triangulated torus.
\item{(iii)}The demonstration that the finite size corrections are modular invariant 
irrespective of the value of the mass and the demonstration that in the massless case
they are related to a topological quantity known as the $\bar\partial$-torsion.
\item{(iv)}The determination of equivalent data for the two dimensional Ising model
on a trinangular lattice with three independent nearest neighbour couplings. 
\par
The organisation of the material in the rest of this paper is as follows. In section two we
discuss our one dimensional theory in its continuum and in its discrete formulation. Section
three moves on to two dimensions where we formulate a gauge theory based on 
flat bundles over the torus $T^2$ and then construct the associated lattice model. 
This section contains some of our basic
results on modular invariance both at and near the critical phase.  In section four we deal with
bulk properties the lattice model including those due to finite size and finite lattice spacing. We also
examine the conformal properties of the model. Finally section  five contains our concluding
remarks.  
\beginsection{A twisted lattice theory in one dimension}
We start in one dimension with a gauge theory on a circle $S^1$ of circumference $L$. Any 
connection $A$ of such a theory necessarily has zero curvature; nevertheless $A$ 
can still be non-trivial since it can have non-trivial holonomy. 
The action $S$ of our theory is given by
$$S={1\over2}\int_{S^1} (\bar f d^*_Ed_E f+m^2\bar ff)\no$$
where (as mentioned in the introduction) $f$ is  {\it not a function} on $S^1$ 
but  a {\it section} of a flat bundle $E$ 
over $S^1$, finally $d_E$ is the covariant derivative which acts on
such sections. 
\par
A few details about the flat bundle $E$: The sections of $E$ must have
non-trivial holonomy and, for simplicity in the subsequent discretisation
of the theory, we take the holonomy group to be ${\bf U}(1)$.
Now a general, rank $n$, vector bundle $E$ over $S^1$ is a
sum of appropriate powers of {\it line } bundles ${\cal L}_i$ i.e. 
$$E={\cal L}_1^{\alpha_1}\oplus {\cal L}_2^{\alpha_2}\oplus\cdots\oplus
{\cal L}_n^{\alpha_n},\quad\hbox{with }\alpha_i\in{\bf Z}\no$$
where ${\cal L}_i$ is a single line bundle over $S^1$.
\par
Because a completely general $E$ is expressible in terms of line bundles ${\cal L}_i$ we shall
just treat the special case where $E$ is a single line bundle and take
$E={\cal L}$ a section $f$ of which has the property that 
$$f(x+nL)=e^{2\pi i n u}f(x).\no$$
The computation of the  partition function for a general $E$ 
(of rank $n$) is an easy extension of the line bundle case. 
\par
Now the partition function $Z(S^1,{\cal L},m)$ for the action $S$  above is given by the
functional integral
$$Z(S^1,{\cal L},m)=\int {\cal D}f {\cal D}\bar f \exp\left[-{1\over2}\int_{S^1} \bar f (d^*_{\cal L}d_{\cal L}+m^2) f\right]\no$$
This is immediately calculable (cf. \ref{6} and \ref{7} and
references therein) and is given by  
$$Z(S^1,{\cal L},m)=\left\{\det[{d_{\cal L}^* d_{\cal L}+m^2\over \mu^2}]\right\}^{-1}\no$$
where $\mu$ is a mass scale introduced to render the partition function dimensionless.
In the above equation $\det[{d_{\cal L}^* d_{\cal L}+m^2\over \mu^2}]$ is given a meaning 
by defining it through the $\zeta$-function regularization procedure.
To implement this procedure we need to find the eigenvalues of the operator 
$d_{\cal L}^* d_{\cal L}+m^2$. Eigensections replace eigenfunctions and a basis of
eigensections is given by 
$$e_n(x)= {1\over\sqrt{L}}\exp^{2\pi i{(n+u)\over L}x}\no$$
Now if we define the inner product between sections by
$$<f,g>=\int_0^Ldx\bar f(x) g(x)\no$$
The relation $<d_{\cal L}f,g>=<f,d^*_{\cal L}g>$ defines the action of $d^*_{\cal L}$
and we find the eigenvalues are 
$$\lambda_n=({2\pi (n+u)\over L})^2+m^2 \no$$
The $\zeta$-function of interest to us is then
$$\zeta_{d_{\cal }^*d_{\cal L}+m^2}(s)={({\mu L\over 2\pi})}^{2s}\sum_{n=-\infty}^{\infty}{1\over {((n+u)^2+({m L\over 2\pi})^2)}^s}.\no$$
The easiest way to evaluate the sum is to use the {\it Plana sum formula} \ref{8}
which can be cast in the form
\eqlabel{\planasum}
$$\eqalign{\sum_{n=-\infty}^{\infty}{1\over {\left((n+u)^2+x\right)}^s}
&={\pi^{1/2}\Gamma(s-{1\over2})\over\Gamma(s)} x^{-s+{1\over2}}\cr
&\qquad\qquad+{\sin(\pi s)\over\pi}\int_{0}^{\infty}dq 
q^{-2s}{d\over dq} \ln\left\vert1-e^{-2\pi\sqrt{q^2+x}+2\pi i u}\right\vert^2
\cr}\no$$
From which we deduce that
$$\zeta_{d_{\cal L}^* d_{\cal L}+m^2}(0)=0\no$$ 
and 
$$\zeta^\prime_{d^*_{\cal L}d_{\cal L}+m^2}(0)=
-mL-\ln\left\vert1-e^{-mL+2\pi i u}\right\vert^2.\no$$
Notice that the arbitrary undetermined scale $\mu$ has completely 
disappeared from the answer and we 
are led to the surprisingly elegant result that the 
$$\det[{d_{\cal L}^* d_{\cal L}+m^2\over \mu^2}]=
\left\vert1-e^{-mL+2\pi i u}\right\vert^2 e^{mL}.\no$$
and hence $\Gamma=-\ln Z(S^1,{\cal L},m)$ is given by
$$\Gamma=mL+\ln\left\vert1-e^{-mL+2\pi i u}\right\vert^2\no$$
In the massless limit, $m=0$, we find that $Z(S^1,{\cal L},0)$ 
is {\it independent} of the circumference $L$ of $S^1$, a reflection of 
metric independence.  In fact the resulting expression is a topological invariant
of the bundle 
$$Z(S^1,{\cal L},0)={1\over T(S^1,{\cal L})}\no$$
where $T(S^1,L)$ denotes the Ray-Singer analytic
torsion\footnote{$^{(a)}$}{\eightpoint 
The equality of the combinatorial and the analytic torsion for
${\cal L}$ provides one with a topological derivation of  Lerch's
classical formula:
$\zeta^\prime_u(0)=-2\ln4-4\ln\sin(\pi u)$, where 
$\zeta_u(s)=\sum_{n=-\infty}^\infty 2/(n+u)^{2s}$. In fact this is
pointed out in Cheeger's paper \ref{9} 
though
the details are different: Cheeger considers, not a (complex) line 
bundle ${\cal L}$, but a (real) $SO(2)$ bundle $E$.} 
of the line bundle ${\cal L}$ and is given by 
$$T(S^1,{\cal L})=\{2\sin(\pi u)\}^2.$$
\par
Though the resulting expressions above are elegant many aspects of the 
procedure that leads to them are mysterious. For example, from 
general principles one would expect that
the partition function would have a term proportional to the number of degrees of
freedom, which would give rise to a divergent contribution in the continuum limit.
It is therefore illuminating to examine the above theory from a lattice point of 
view. We therefore turn to its discretisation. 
\par
To this end we replace the circle $S^1$ with its natural combinatorial
counterpart namely a polygon with $N$ vertices. The discretised basis 
of eigensections is then take to be
$$\{{1\over\sqrt{Na}}e^{i x_n k}:\;k=1,\dots,N\}
\quad\hbox{where }x_n={2\pi (n+u)\over N},\;n=0,\ldots,N-1.\no$$ 
where the lattice spacing $a={L\over N}$. 
Since the connection is locally trivial the action of the discretised derivative 
$d_{\cal L}$ on sections can be taken to be 
$$d_{\cal L}f(k)={f(k+1)-f(k)\over a} \no $$
The discretised inner product becomes
$$<f,g>=a\sum_{k=0}^{(N-1)}\bar f(k) g(k)\no$$
and hence 
$$d^*_{\cal L}f(k)={f(k-1)-f(k)\over a}.\no$$
Applying these rules to the basis $e_n(k)$ implies that
$$\eqalign{d_{\cal L}e_n(k)&={2i\over a}e^{i{x_n\over2} }\sin\left({x_n\over2}\right)e_n(k)\cr
\hbox{and }\quad d^*_{\cal L}e_n(k)&=-{2i\over a}e^{-i{x_n\over 2}}
\sin\left({x_n\over2}\right)e_n(k)                    \cr}\no$$
and so we find that the finite dimensional lattice version of 
$\det[d^*_{\cal L} d_{\cal L}+m^2]$, which we shall denote by 
$\det_N[d^*_{\cal L} d_{\cal L}+m^2]$, is given by
$${\det}_N[d^*_{\cal L} d_{\cal L}+m^2]=\prod_n\lambda_n,\quad
\hbox{where }
\lambda_n={1\over a^2}\sin(x_n)^2+m^2\no$$
The corresponding lattice partition function  we denote by $Z_N(S^1,{\cal L},m)$, and it is, 
of course, automatically finite.  We have
$$\eqalign{Z_N(S^1,{\cal L})&=\int\prod_{k}d\phi_1(k)d\phi_2(k)\exp[-S]\qquad
\hbox{where }\quad S={1\over2}<d_{\cal L}\varphi,d_{\cal L}\varphi>.\cr}\no$$
We have taken $\phi=\phi_1+i\phi_2$ and  
$\varphi=\sqrt{a}\phi$, then $\varphi$ corresponds to the usual continuum field 
and $\phi$ is a dimensionless lattice field. 
This identification renders the partition function dimensionless as we require it to 
be and in fact we have
$$Z_N(S^1,{\cal L},m)=\left\{{\det}_N{(d^*_{\cal L}d_{\cal L}+m^2)a^2\over 2\pi}\right\}^{-1}\no$$
Defining $W=-\ln{Z_N}$ we find that $W$ is given by the sum
$$\eqalign{W=&\sum_{n=0}^{N-1}\ln{{\lambda_na^2\over 2\pi}}\cr
=&-N\ln\pi+\sum_{n=0}^{N-1}\ln[1+{m^2a^2\over2}-\cos(x_n)]}\no$$
The sun can be performed using the basic identity
$$\sum_{n=0}^{N-1}\ln[z-\cos(x_n)]=N\ln[{z_+\over2}]+\ln\vert1-z_-^Ne^{2\pi i u}\vert^2\no$$
where $z_{\pm}=z\pm\sqrt{z^2-1}$. 
Hence we find that $W$ splits in the form
$$W=NW_B+W_F$$
where the extensive term is known as the bulk contribution. The quantity 
$W_B$ is the contribution to $W$ per lattice site in the thermodynamic limit i.e.
$$W_B=\lim_{N\rightarrow\infty}{W\over N}$$
and is given by
$$W_B=\ln[1+{m^2a^2\over2}+\sqrt{m^2a^2+m^4a^4\over4}]-\ln2\pi.\no$$
The remaining term $W_F$  captures the effects of a finite lattice
and is given by
$$W_F=\ln\vert1-(1+{m^2a^2\over2}-\sqrt{m^2a^2+{m^4a^4\over4}})^Ne^{2\pi i u}\vert^2\no$$

The continuum or scaling limit is a constrained thermodynamic limit and 
corresponds to taking $N\rightarrow\infty$ while keeping
$Na=L$ fixed. In this limit we have 
$$NW_B=-N\ln2\pi+\Gamma_B\qquad\hbox{where}\qquad\Gamma_B=mL\no$$ 
and 
$$\Gamma_F=\ln\vert1-e^{-mL+2\pi iu}\vert^2\no$$
We see therefore that the quantity 
$W+N\ln2\pi$ tends to a finite continuum limit which in fact coincides
with the result provided by the $\zeta$-function prescription for the 
functional determinant.  

It is clear from the above that we could alter the integration measure to absorb the 
term $-N\ln 2\pi$. This is most naturally done by including the Van Vleck-Morette 
determinant $J$ in the integration measure so that 
$$\eqalign{Z^c_N(S^1,{\cal L},m)&=\int\prod_{k}d\phi_1(k)d\phi_2(k)J\exp[-S]\cr
\qquad
\hbox{where }\quad J&=\left\{\det\left({1\over2\pi}
{\partial^2S\over\partial\phi_i(k)\partial\phi_{j}(k-1)}\right)\right\}^{1/2}.\cr}\no$$
The resulting partition function $Z^c_N(S^1,{\cal L},m)$  is given by
$$\eqalign{Z^c_N(S^1,{\cal L},m)&=z_+^N\left\vert1-z_-^Ne^{2\pi i u}\right\vert^2\cr
\hbox{where}\qquad z_{\pm}&=1+{m^2a^2\over2}\pm\sqrt{m^2a^2+{m^4a^4\over4}}}\no$$
With $L=Na$ fixed we have 
$$\lim_{N\rightarrow\infty} Z^c_N(S^1,{\cal L},m)=Z(S^1,{\cal L},m)\no$$
But notice that if $m=0$ then 
$$Z^c_N(S^1,{\cal L},0)={1\over T(S^1,{\cal L})}\no$$ 
In other words the discrete partition function with the Van Vleck-Morette determinant 
included in the continuum limit yields exactly the expression obtained by the 
$\zeta$-function method; in addition in the massless limit this lattice partition function
$Z^c_N(S^1,{\cal L},0)$
coincide exactly with  $Z(S^1,{\cal L},0)$ and is therefore inversely proportional to 
the torsion {\it without the need for a limit.}  The partition function $Z^c_N(S^2,{\cal L},0)$ is 
therefore closely related to the combinatorial torsion and can be viewed as providing a realization
of this object. 
\par
\beginsection{A two dimensional theory at genus one}
Now we go to two dimensions where we want to study a  
field theory on a Riemann surface of genus one i.e. a torus which we
denote by $T^2$.  We shall then go on to discretise with a specific
triangulation to be described below. We begin by 
we providing a brief summary of the continuum setting.
\par
First of all let us realise the continuum torus $T^2$ as a complex
manifold using the standard quotient
\eqlabel{\quotientaction}
$$T^2={\bf C}/\Gamma\no$$
$$\qquad\hbox{\rm with }
\Gamma={\bf Z}\oplus{\bf Z}
               \simeq\pi_1(T^2)\no$$
Next, because we are going to have flat connections, 
we need a flat bundle $E$ over $T^2$, this is obtained by choosing an
element $\chi$ of
$${\rm Hom}\,(\pi_1(T^2),G)\no$$
where $G$ is the bundle structure group. We take $E$ to be a line bundle
${\cal L}$ so that $G=U(1)$. Then $\chi\in{\rm Hom}\,(\pi_1(T^2),G)$ is of the
form
$$\chi:\pi_1(T^2)\longrightarrow S^1\no$$
and $\chi$ is then actually a character  of $\pi_1(T^2)$ which is the reason
for adopting the present notation.
\par
The action we choose is 
 given by
$$S={1\over2}\int_{T^2}\,\bar\nu (d^*_{\cal L}d_{\cal L}+m^2) \nu\no$$
where $m$ is a mass.
Now the torus is a K\"ahler manifold and so has the property
that $d^*_{\cal L}d_{\cal L}=2\bar\partial_{\cal L}^*\bar\partial_{\cal
L}$, where $\bar\partial=\partial/\partial \bar z$.  
This means that the partition function $Z(T^2,{\cal L},m)$ for this action 
gives the $\bar\partial$-torsion if we set $m=0$, i.e.
\eqlabel{\torusdetrelns}
$$\eqalign{Z(T^2,{\cal L},0)&=\int {\cal D}\nu {\cal D}\bar\nu 
\exp\left[-{1\over2}\int_{T^2}\bar \nu {d_{\cal L}}^*d_{\cal L}\nu\right]\cr
                   &=\left\{\det({d_{\cal L}^*d_{\cal L}\over\mu^2})\right\}^{-1}\cr
                   &=\{T_{\bar\partial}(T^2,{\cal L})\}^{-1}\cr}            \no$$
where $T_{\bar\partial}(T^2,{\cal L})$ denotes the 
$\bar\partial$-torsion\footnote{$^{(b)}$}{\eightpoint The 
 $\bar\partial$-torsion is the complex analogue of the ordinary
Ray-Singer analytic torsion $T(M,E)$. For even dimensional $M$ the ordinary 
torsion is always unity; however if the manifold
$M$ is complex then there exists the analytic $\bar\partial$-torsion which is
non-trivial cf. \ref{10}.} and  $\det(\bar\partial_{\cal L}^*\bar\partial_{\cal L})$
is calculated (as is the analytic $\bar\partial$-torsion) using a zeta
function and it is useful to note in the present context that
$$\eqalign{d^*_{\cal L}d_{\cal L}&=2\bar\partial_{\cal L}^*\bar\partial_{\cal  L}\cr
           \Rightarrow \zeta^\prime_{d^*_{\cal L}d_{\cal
L}}(0)&=\zeta^\prime_{\bar\partial_{\cal L}^*\bar\partial_{\cal  L}}(0)
-\ln(2)\zeta_{{\bar\partial_{\cal L}^*\bar\partial_{\cal  L}}(0)}(0)\cr
&=\zeta^\prime_{\bar\partial_{\cal L}^*\bar\partial_{\cal 
L}}(0),\quad\hbox{since calculation will show that }
\zeta_{{\bar\partial_{\cal L}^*\bar\partial_{\cal  L}}(0)}(0)=0\cr
\Rightarrow \det(d^*_{\cal L}d_{\cal L})&=\det(\bar\partial_{\cal L}^*\bar\partial_{\cal L})\cr}\no$$
\par
Now the $\bar\partial$-torsion $T_{\bar\partial}(T^2,{\cal L})$  is a rather
complicated object and was shown in \ref{10} to be expressible in terms
of a ratio containing a Jacobi theta function and the Dedekind eta
function. Before giving $T_{\bar\partial}(T^2,{\cal L})$ we need to give a
parametrisation  of the homomorphism $\chi$ and to do this we use a
notation similar, but not identical, to that of \ref{10}: Let the  torus $T^2$ be
denoted in standard
fashion by the usual parameter $\tau\in {\bf C}^+$ 
(${\bf C}^+$ denotes the upper complex half-plane $\im z>0$), the
generators of $\pi_1(T^2)$ are denoted by the pair $\{1,\tau\}$ and an
element of $\pi_1(T^2)$ by the pair of integers $(m,n)$. Then if $u_0,u_1\in[0,1]$
one has
\eqlabel{\chidef}
$$\eqalign{\chi:&\,\pi_1(T^2)\longrightarrow S^1\cr
&(m,n)\longmapsto \exp[2\pi i(mu_0+nu_1)]\cr}\no$$
While for the  $T_{\bar\partial}(T^2,{\cal L})$  we have 
\eqlabel{\dbartorsion}
$$T_{\bar\partial}(T^2,{\cal L})=
\left\vert\exp[\pi i u_0^2\tau]{\theta_1(u_1-\tau u_0,\tau)\over\eta(\tau)}\right\vert^2\no $$
\par
We finish this continuum summary with a  description of
the continuum spectral
data for $\bar\partial_{\cal L}^*\bar\partial_{\cal L}$: For simplicity we shall work with the
Laplacian $d_{\cal L}^*d_{\cal L}$ and denote it 
by $\Delta_{\cal L}$; its  eigensections by ${\cal E}_{mn}(x,y)$ and eigenvalues
by $\lambda_{mn}$ can be computed fairly straightforwardly.  The details
are as follows: let $z=x+iy$ denote the local coordinate in ${\bf C}$ then
take the torus to have periods $L_0$ and $L_1$ but at an angle $\theta$ to one another.
Then a point $z$ and $z+ (m\tau+n)L_0$ represents points that are identified under the quotient action (\docref{quotientaction}) where 
$$\tau=\tau_0+i\tau_1 \qquad\hbox{with}\qquad\tau_0={L_1\over L_0}\cos\theta\qquad\hbox{and}\qquad\tau_1={L_1\over L_0}\sin\theta.\no$$
It is sensible to change $(x,y)$ to
new (real) variables $(x_0,x_1)$ defined by writing
$$z=x_0+x_1\cos\theta +i x_1\sin\theta.$$
In these new co-ordinates the metric $ds^2=dx^2+dy^2$ becomes
$ds^2=dx_0^2+2\cos\theta dx_0dx_1+dx_1^2$ and the 
Laplacian $\Delta_{\cal L}$ becomes 
$$\Delta_{\cal L}=-{1\over\sin^2\theta}\left({\partial^2\over\partial x_0^2}-\cos\theta{\partial^2\over\partial x_0\partial x_1}+{\partial^2\over\partial x_1^2}\right)\no$$
It is then easy to find that the eigensections are now given by
$${\cal E}_{mn}(x_0,x_1)=\exp\left[2\pi i\{{(n_0+u_0)\over L_0}x_0
+{(n_1+u_1)\over L_1}x_1\}\right]\no$$
Note that this means that the ${\cal E}_{mn}(x_0,x_1)$ are not periodic,
being sections of ${\cal L}$, but have the holonomy predictable from the character 
$\chi$ defined in \docref{chidef} above. The eigenvalues can be easily found and are given 
by 
$$\eqalign{\lambda_{mn}&=\left({2\pi\over L_0\tau_1}\right)^2\left[\vert\tau\vert^2(n_0+u_0)^2-2(n_0+u_0)(n_1+u_1)\tau_0+(n_1+u_1)^2\right]\cr
&=\left({2\pi\over L_0\tau_1}\right)^2\left\vert(n+u_1)-\tau(m+u_0)\right\vert^2}\no$$
A computation of $\det\Delta_{\cal L}$ by the $\zeta$-function method establishes 
(\docref{torusdetrelns}) where  $T_{\bar\partial}(T^2,{\cal L})$ is given by (\docref{dbartorsion}).
\par
This completes our brief description of the continuum spectral data 
and we pass on to the matter of discretisation where we shall be able to
reproduce exactly the continuum partition function---the inverse of the 
$\bar\partial$-torsion---in calculating the finite size effects on the lattice.
\par
To this end we now  replace the $\tau$ parallelogram by a 
discrete  lattice which we then triangulate as follows: To retain
complete generality over the geometry of the triangulation we construct 
a triangular lattice composed of similar triangles, pairs of which form parallelograms, 
The basic triangles have two sides of lengths $a_0$ and $a_1$ with an angle $\theta$ between
them. 
The complete lattice forms our torus, $T^2$, 
and consists of $K_0K_1$ sites and $2K_0K_1$ triangles, 
forming a parallelogram of sides $L_0=K_0a_0$ and $L_1=K_1a_1$.
\par
The resulting lattice is depicted in fig. 1 and we can use this figure
to deduce that with this model for the
torus a point $z\in{\bf C}$ is identified with points of the form 
$z+m L_0+n L_1 \exp[i\theta]$ where $m,n\in{\bf Z}$. Hence if $z=(x,y)$ then
$$(x,y)\equiv (x+m L_1\cos\theta+n L_0,y+m L_1\sin\theta),\quad m,n\in{\bf Z}\no$$ 
This geometric information also determines the holonomy 
of a section $\varphi$ of the  bundle ${\cal L}$
over $T^2$ which we record in the equation 
\eqlabel{\bdryconds}
$$\varphi(x+mL_1\cos\theta+nL_0,y+mL_1\sin\theta)=e^{2\pi
i(mu_1+nu_0)}\varphi(x,y),\quad m,n\in{\bf Z}\no$$
\par
We shall take a statistical mechanical standpoint and compute the
energy of the lattice configuration rather than the, equally possible,
approach of computing its action. To do this, we label the lattice sites by
$(k_0,k_1)\equiv{\bf k}=k_0+(k_1-1)K_0$
with $k_i=1,\dots,K_i$, and the  energy is then given by
\eqlabel{\quadform}
$$\eqalign{{\cal E}_{\cal L}[T^2,\varphi^*,\varphi]&={1\over2}\sum_{{\bf k}{\bf k}^\prime}
\sqrt{g}\varphi^{*}({\bf k})
\left(\Delta({\bf k},{\bf k}^\prime)
+m^2\delta_{{\bf k},{\bf k^\prime}}\right)\varphi({\bf k}^\prime)\cr
}\no$$ 
where $\sqrt{g}=a_0a_1\sin\theta$. 
\par
The discrete eigensections of $\Delta$ are ${\cal E}_{n_0n_1}(k_0,k_1)$ where 
$${\cal E}_{n_0n_1}(k_0,k_1)=\exp\left[2\pi i\left\{(n_0+u_0){k_0\over
K_0}+(n_1+u_1){k_1\over
K_1}\right\}\right]\no$$
and the discrete version of the Laplacian $\Delta$ is a
$K_0K_1\times K_0K_1$ symmetric matrix $\Delta(K_0,K_1)$ a general element of
which we denote by $\Delta\{(k_0,k_1),(k_0^\prime,k_1^\prime)\}$ so that we have
$$\eqalign{&\Delta(K_0,K_1)=
\left[\Delta\{(k_0,k_1),(k^\prime,l^\prime)\}\right]_{K_0K_1\times K_0K_1}\cr
&\hbox{\rm with } 
1\le k_0,k_0^\prime\le K_0,\quad \hbox{\rm and }
 1\le k_1,k_1^\prime\le  K_1\cr}\no$$ 
Using a nearest neighbour interaction all the non-zero entries of 
$\Delta(K_0,K_1)$ 
can be deduced by its symmetry and by recording explicitly that
$$\eqalign{
\Delta\{(k_0,k_1),(k_0+1,k_1)\}&=-\alpha=-{1\over\sin^2\theta}
\left({1\over a_0^2}-{\cos\theta\over a_0a_1}\right)\cr
\Delta\{(k_0,k_1),(k_0,k_1+1)\}&=-\beta=-{1\over\sin^2\theta}
\left({1\over a_1^2}-{\cos\theta\over a_0a_1}\right)\cr
\Delta\{(k_0+1,k_1),(k_0,k_1+1)\}&=-\gamma=-{1\over\sin^2\theta}\left({\cos\theta\over a_0a_1}\right)
\cr
\Delta\{(k_0,k_1),(k_0,k_1)\}&=2\sigma,\quad\hbox{where
}\sigma=\alpha+\beta+\gamma\cr
\cr}
\no$$
the remaining elements being zero. The lattice eigenvalues are then
\eqlabel{\lateigenvals}
$$\eqalign{\lambda_{n_0n_1}
&=2\left(\sigma-\alpha\cos\left(x_{n_0}\right)-\beta\cos\left(x_{n_1}\right)
-\gamma\cos\left(x_{n_0}-x_{n_1}\right)\right)\cr
 \hbox{where } 
x_{n_i}&={2\pi(n_i+u_i)\over K_i},\;i=0,1\cr}
\no$$
\par
The partition function that interests us is 
$$Z_{K_0K_1}(T^2,{\cal L},m)=\int\left[\prod d\varphi^* d\varphi\right]
e^{-{\cal E}_{\cal L}[T^2,\varphi^*,\varphi]}
=\prod_{(n_0,n_1)}^{(K_0,K_1)}\left\{{2\pi\over
\sqrt{g}(\lambda_{n_0n_1}+m^2)}\right\}
\no$$
and $W=-\ln Z$.
\par
We pass now to the lattice determinant in which we are interested namely
$$\det\left({\Delta({\bf k},{\bf k}^\prime)
+m^2\delta_{{\bf k},{\bf
k^\prime}}\over2}\right)=\prod_{(n_0,n_1)}\left({\lambda_{n_0n_1}+m^2\over2}\right)\no$$
The computation of this determinant is a non-trivial task but one that
we shall accomplish. 
\par
Purely for convenience of the reader following the calculation 
we set $K_0=K_1\equiv K$---all our results can be derived 
without making this specialisation and we shall quote formulae
later (in the next section)  without this restriction imposed.
Now we want to compute $\prod_{(n_0,n_1)}(\lambda_{n_0n_1}+m^2)/2$ 
and it suits us to compute the logarithm of this quantity
which we shall denote by $\ln\det{}_K((\Delta+m^2)/2) $ for short, i.e.
\eqlabel{\detdef}
$$\ln\det{}_K\left({\Delta+m^2\over2}\right)
=\sum_{n_0,n_1=0}^{(K-1)}\ln\left({\lambda_{n_0n_1}+m^2\over2}\right)\no$$
It turns out that we can actually make sense of this expression and even
evaluate its $K\rightarrow\infty$ continuum limit where we shall find a
{\it modular invariant} expression. A key step in achieving this is that we do
one of the sums in \docref{detdef} completely. We now outline how this
latter step comes about.
\par
Let us define $\delta$, by $\delta=\sigma+m^2/2$, so that using \docref{lateigenvals}
this implies that
$$\lambda_{n_0n_1}+m^2=2\left(\delta-\alpha\cos\left(x_{n_0}\right)-\beta\cos\left(x_{n_1}\right)
-\gamma\cos\left(x_{n_0}-x_{n_1}\right)\right)\no$$
Define the new quantities, $b_{n_1}$, $z_{n_1}$ and $\theta_{n_1}$ by writing
$$\eqalign{b_{n_1}=(\alpha+\gamma\exp[-ix_{n_1}])\equiv\vert
b_{n_1}\vert\exp[i\theta_{n_1}],\quad
z_{n_1}={(\delta-\beta\cos(x_{n_1}))\over\vert b_{n_1}\vert}\cr
\Rightarrow \left[\delta-\alpha
\cos(x_{n_0})-\beta\cos(x_{n_1})-\gamma
\cos(x_{n_0}-x_{n_1})\right]=\vert b_{n_1}\vert(z_{n_1}-\cos(x_{n_0}+\theta_{n_1}))\cr
}\no$$
Then we {\it factorise} $(z_{n_1}-\cos(x_{n_0}+\theta_{n_1}))$ by observing that
\eqlabel{\factorformula}
$$\eqalign{(z_{n_1}-\cos(x_{n_0}+\theta_{n_1}))&={z_+\over2}(1-z_-\exp[i(x_{n_0}+\theta_{n_1})])(1-z_-\exp[-i(x_{n_0}+\theta_{n_1})])\cr
\hbox{where }z_{\mp}=z_{n_1}\mp\sqrt{z_{n_1}^2-1}\cr}\no$$
Hence the sum over $n$ in \docref{detdef} requires us to compute
$$\eqalign{\ln\det{}_K\left({\Delta+m^2\over2}\right)
&=\sum_{n_0,n_1=0}^{(K-1)}\ln\left({\lambda_{n_0n_1}+m^2\over2}\right)\cr
                      &=\sum_{n_0,n_1=0}^{(K-1)}\ln\vert
                        b_{n_1}\vert(z_{n_1}-\cos(x_{n_0}+\theta_{n_1}))\cr
&=K\sum_{n_1=0}^{(K-1)}\ln({z_+\vert b_{n_1}\vert\over 2})+
\sum_{n_0,n_1=0}^{(K-1)}\ln\vert 1-z_-\exp[i(x_{n_0}+\theta_{n_1})]\vert^2}\no$$

But 
\eqlabel{\trick}
$$\eqalign{\sum_{n_0=0}^{(K-1)}\ln(1-z_-\exp[i(x_{n_0}+\theta_{n_1})])&=
-\sum_{n_0=0}^{(K-1)}\sum_{r=1}^{\infty}{(z_-)^r\over r}\exp[ir(x_{n_0}+\theta_{n_1})]\cr
&=-\sum_{r=1}^{\infty}{(z_-\exp[i\theta_{n_1}])^r\over
r}\sum_{n_0=0}^{(K-1)}\exp[ir x_{n_0}]\cr
&=-\sum_{l=1}^{\infty}{(z_-\exp[i\theta_{n_1}])^{lK}\over lK}K\exp[{2\pi
i u_0 l}]\cr
&=\ln(1-v_{n_1}^K)\cr
\hbox{where }v_{n_1}&=z_-\exp[i\theta_{n_1}]\exp[2\pi i u_0/K]\cr}\no$$
 Now\footnote{$^{(c)}$}{\eightpoint In \docref{trick} above we use the fact that $\exp[2\pi i r/K]$ are
the $K^{th}$ roots of unity so that 
$$\sum_{r=0}^{(K-1)}\exp[2\pi i nr/K]=
                                       \cases{K, &if $n=0 \,\mod K$\cr
                                              0, &otherwise\cr}\no$$
 a similar algebraic trick was used in \ref{11}.}
 we note that we have entirely done one of the summations. It turns out
that we can do still more: for the first term
$\ln\left({z_+\vert b_{n_1}\vert/2}\right)$ we can accomplish
the remaining sum over $n_1$. We have
$$\eqalign{\ln\left({z_+\vert b_{n_1}
\vert\over 2}\right)&=
\ln\left[{1\over 2}\left({A_{n_1}\over2}+\sqrt{{A_{n_1}^2\over4}-\vert
b_{n_1}\vert^2}\right)\right]\cr
\hbox{ where }A_{n_1}&=2\delta-2\beta\cos(x_{n_1})\cr}\no$$
But using the integral identity
$$\int_0^{\pi/2}{d\omega \over\pi} \,
\ln(z^2-\lambda^2\cos^2(\omega))=\ln\left({z+\sqrt{z^2-\lambda^2}\over2}\right)\no$$
we find that
$$\ln\left({z_+\vert b_{n_1}
\vert\over
2}\right)=\int_0^{\pi/2}{d\omega \over\pi} \,
\ln\left[\left({A_{n_1}\over2}^2-\vert
b_{n_1}\vert^2\cos^2(\omega)\right)\right]\no$$
Now we have to do the summation and evaluate
\eqlabel{\tempexp}
$$\sum_{n_1=0}^{(K-1)}K\ln\left({z_+\vert b_{n_1}
\vert\over 2}\right)=\sum_{n_1=0}^{(K-1)}
K\int_0^{\pi/2}{d\omega \over\pi} \,
\ln\left[\left({A_{n_1}\over 4}^2-\vert
b_{n_1}\vert^2\cos^2(\omega)\right)\right]
\no$$
Next we reuse the two properties summarised in
\docref{factorformula} and \docref{trick} to do the summation. But first we
must rewrite \docref{tempexp} in a suitable factorised form: noting that
$A_{n_1}$ is quadratic in $\cos(x_{n_1})$ we get
$$\eqalign{{A_{n_1}^2\over 4}-\vert b_{n_1}
\vert^2\cos^2(\omega)&=\beta^2(\cos^2(x_{n_1})-2q\cos(x_{n_1})+2a)\cr
                &=\beta^2(\cos(x_{n_1})-x^0)(\cos(x_{n_1})-x^1)\cr
 \hbox{ with }&\cases{x^0=q+\sqrt{q^2-a^2}\cr
                     x^1=q-\sqrt{q^2-a^2}\cr}\cr  
\hbox{So \docref{factorformula}}\Rightarrow      
{A_{n_1}^2\over 4}-\vert b_{n_1} \vert^2\cos^2(\omega)&=\beta^2
{x^0_+\over2}{x^1_+\over2}\vert 1-x^0_-\vert^2\vert 1-x^1_-\vert^2\cr 
\hbox{ where }&\cases{x^0_\mp=x^0\mp\sqrt{(x^0)^2-1}\cr
                      x^1_\mp=x^1\mp\sqrt{(x^1)^2-1}\cr}
                              \cr}\no$$
If we now employ \docref{trick} we find that we can do the sum over $n_1$
leaving us with a trivial two term sum originating in the roots of the 
quadratic. We get 
$$\eqalign{\sum_{n_1=0}^{(K-1)}K\ln\left({z_+\vert b_{n_1}
\vert\over 2}\right)&=K
\sum_{i=0}^{1}\left\{
\int_0^{\pi/2}{d\omega \over\pi} \,K\ln\left({\beta
x^i_+\over 2}\right)
+\ln\vert
1-(s^i_-)^K\vert^2\right\}\cr
\hbox{ where }s^i_-&=x^i_-\exp\left[{2\pi i u_1\over K }\right],\;i=0,1\cr}\no$$
\par
Thus we have accomplished our summation goal and have found for
$\ln\det{}_K\left({\Delta+m^2\over2}\right) $ 
the expression
\eqlabel{\mainexpress}
$$\eqalign{\ln\det{}_K\left({\Delta+m^2\over2}\right)=
K^2\int_0^{\pi/2}{d\omega\over\pi}\,
\ln\left({\beta^2x^0_+x^1_+\over 4}\right)&+
K\sum_{i=0}^1 \int_0^{\pi/2}{d\omega\over\pi}\,
\ln\vert
1-(s^i_-)^K\vert^2\cr
&+\sum_{n_1=0}^{(K-1)}\ln\vert 1-v_{n_1}^K\vert^2\cr}\no$$ 
Actually the coefficient of $K^2$  above (the bulk coefficient), 
is easily checked to be given by 
$$\int_0^{\pi/2}{d\omega\over\pi}\,
\ln\left({\beta^2x^0_+x^1_+\over 4}\right)=
\int_{-\pi}^{\pi}{d\nu_1d\nu_2\over (2\pi)^2}
\ln\left[\left\{
\delta-\alpha\cos(\nu_1)-\beta\cos(\nu_2)-\gamma\cos(\nu_1-\nu_2)
\right\}\right]\no$$
\par
We can now
begin to 
calculate what happens when $K$ is large. 
{\it Note that we shall now  (temporarily) 
specialise to the case where  $m=0$ since 
our answer in this case should be checkable 
against the 
$\bar\partial$-torsion  
$T_{\bar\partial}(T^2,{\cal L})$ }. However 
after our successful check 
against  $T_{\bar\partial}(T^2,{\cal L})$   
we shall give the result for $m\not=0$ also.
We shall start with the two terms $\ln\vert
1-(s^i_-)^K\vert^2$, $i=0,1$ it turns out that 
\eqlabel{\twolimits}
$$\lim_{K\rightarrow\infty}K\int_0^{\pi/2}{d\omega\over\pi}\,
\ln\vert
1-(s^i_-)^K\vert^2=\cases{0,& for $i=0$ because $\vert s_-^0|<1$ for
                                                     $\omega\in [0,\pi/2]$\cr
                          \not =0,& for $i=1$ because $\vert s_-^1|\le 1$
                          for
                                                     $\omega\in [0,\pi/2]$\cr}\no$$
\par
We briefly describe the calculation of the surviving limit in
\docref{twolimits} above.  The entire contribution to the limit comes
from the $\omega=0$ end of the integral since it is only at $\omega=0$
that $\vert s_1\vert$ attains the value $1$. Taking this as our starting
point we just need  the (laboriously obtained) fact that, near
$\omega=0$, 
$$x^1_-=1-{\vert \tau\vert^2\over\tau_1}\omega+\cdots\no$$
which is relevant  since $s^1_-=x^1_-\exp[2\pi i u_1/K]$. This found we
write
$$\int_0^{\pi/2}{d\omega\over\pi}\,
\ln\vert
1-(s^1_-)^K\vert^2=\int_0^\epsilon{d\omega\over\pi}\,
\ln\vert
1-(s^1_-)^K\vert^2+\int_\epsilon^{\pi/2}{d\omega\over\pi}\,
\ln\vert
1-(s^1_-)^K\vert^2,\;\epsilon>0\no$$
The second integral dies exponentially as $K\rightarrow\infty$ but
the first integral gives, for small $\epsilon$,

$$\eqalign{\int_0^\epsilon{d\omega\over\pi}\,
\ln\vert
1-(s^1_-)^K\vert^2&=-\sum_{l=1}^\infty
\int_0^\epsilon{d\omega\over\pi}\,(s^1_-)^{Kl}\cr
\Rightarrow K\int_0^\epsilon{d\omega\over\pi}\,
\ln\vert
1-(s^1_-)^K\vert^2&\rightarrow
-K\sum_{l=1}^\infty\int_0^\epsilon{d\omega\over\pi}\,
{\exp[2\pi i u_1
l]\over l}\left(1-{\vert \tau\vert^2\over\tau_1}\omega\right)^{Kl}\cr
&=-K\sum_{l=1}^\infty{\exp[2\pi i u_1 l]\over l\pi}
\left({-\tau_1\over\vert\tau\vert^2(Kl+1)}\right)
\left[\left(1-{\vert
\tau\vert^2\over\tau_1}\omega\right)^{Kl+1}\right]^\epsilon_0
\cr}\no$$
From which we readily compute that
$$\eqalign{\lim_{K\rightarrow\infty}K\int_0^{\pi/2}{d\omega\over\pi}\,
\ln\vert
1-(s^1_-)^K\vert^2&=-{2\tau_1\over\pi \vert\tau\vert^2}\sum_{l=1}^\infty{\cos(2\pi u_1 l)\over
l^2}\cr
       &=-{2\pi \tau_1\over\vert\tau\vert^2}\left(u_1^2-u_1+{1\over 6}\right)\cr}\no$$
Having taken care of the second term in \docref{mainexpress} we move on. 
\par
The first term in \docref{mainexpress} already has its $K$-dependence displayed as an
explicit factor of $K^2$ and so the only other term to consider is 
the last term , namely $\sum_{n_1=0}^{(K-1)}\ln\vert
1-v_{n_1}^K\vert^2$. We note that, apart
from an explicit factor of $K$, the
dependence on $K$ in the sum is via the variable
$$x_{n_1}={2\pi(n_1+u_1)\over K}\no$$
and for $K\rightarrow\infty$ we distinguish the two cases
$$K\rightarrow\infty\cases{n_1/K\rightarrow0,&case 0\cr
                           n_1/K\not\rightarrow0,&case 1\cr}\no$$
Since the procedure for both cases is lengthy but reasonably
straightforward we describe it fairly briefly: first we deal with {\it
case $0$} and then summarise the differences relevant for  {\it
case $1$}. 
\par\noindent
{\bf Case 0:}
\par
We consider first the term $\ln\vert
1-v_{n_1}^K\vert^2$ 
and simply quote the fact that, after slightly tedious algebra, one
finds that, with
$\lambda=\sqrt{(\alpha\beta+\beta\gamma+\gamma\alpha)}/(\gamma+\alpha)$, 
we have
$$\eqalign{v_{n_1}&\longrightarrow
\exp\left[-{2\pi \lambda(n_1+u_1)\over K}\right]\left(1-{i\gamma\over(\alpha+\gamma)}{2\pi(n_1+u_1)\over K}\right)
\exp\left[{2\pi i u_0\over K}\right]\cr&
=\exp\left[{-2\pi\tau_1\over K\vert\tau\vert^2}\right]
\left(1-{i\tau_0\over\vert\tau\vert^2}{2\pi(n_1+u_1)\over K}\right)\exp\left[{2\pi i u_0\over K}\right]\cr
            }\no$$
Hence we now obtain
\eqlabel{\storeone}
$$\eqalign{\lim_{n_1/K\rightarrow0}v_{n_1}^K&=\exp\left[-2\pi i\left\{(n_1+u_1){\bar
\tau\over\vert\tau\vert^2}-u_0\right\}\right]\cr
&=\exp\left[-2\pi i\left\{{(n_1+u_1)
\over\tau}-u_0\right\}\right]\cr}\no$$
This expression \docref{storeone} will take care of the
$K\rightarrow\infty$
 limit of the  term $\ln\vert
1-v_{n_1}^K\vert^2$. We must now round off the treatment of the
$K\rightarrow\infty$ limit  by describing what happens for {\it case 1}.
\par\noindent
{\bf Case 1:}
\par
Recall that {\it case 1} means that  $n_1/K\not\rightarrow0$ as
$K\rightarrow\infty$. We can achieve this by writing
$$n_1=(K-l),\quad l=1,\dots,(K-1)\no$$
But since $v_{n_1}$ only depends on $n_1$ via the functions $\cos(x_{n_1})$ and
$\sin(x_{n_1})$
it is enough to note that with $n_1\mapsto(K-l)$ we obtain
$$\eqalign{\cos(2\pi(n_1+u_1)/K)\equiv\cos(x_{n_1})&\mapsto \cos(2\pi(l-u_1)/K)\cr
           \sin(2\pi(n_1+u_1)/K)\equiv\sin(x_{n_1})&\mapsto
-\sin(2\pi(l-u_1)/K)\cr}\no$$
The discussions of case 0 and case 1 are now complete  and this allows us to 
deduce that
$$\lim_{K\rightarrow\infty,\,n_1/K\not\rightarrow0
          }v_{n_1}^K=\exp\left[2\pi i\left\{{(l-u_1)
\over\bar\tau}+u_0\right\}\right]\no$$
We now have the entire $K\rightarrow\infty$ limit of $\ln\det{}_K\left(\Delta/2\right) $ and
it is that, as $K\rightarrow\infty$ 
$$\eqalign{\ln\det{}_K&\left({\Delta\over2}
\right) \longrightarrow K^2
\int_{-\pi}^{\pi}{d\nu_1d\nu_2\over (2\pi)^2}
\ln\left[\left\{
\sigma-\alpha\cos(\nu_1)-\beta\cos(\nu_2)-\gamma\cos(\nu_1-\nu_2)
\right\}\right]\cr
&-{2\pi \tau_1\over\vert\tau\vert^2}\left(u_1^2-u_1+{1\over 6}\right)
+\sum_{n=-\infty}^{\infty}\ln\left\vert 1-\exp\left[-2\pi i\left\{{\vert
n\vert\over \tau}+\epsilon_n\left({u_1\over\tau}-u_0\right)\right\}\right]
 \right\vert^2\cr
&\hbox{where }\epsilon_n=\cases{1,& $n\ge$ 0\cr -1,& $n<0$\cr}\cr}\no$$
Hence the finite part of $\ln\det{}_K\Delta_{\cal L}/2 $ is the function $F(u_0,u_1,\tau)$ given by
$$F(u_0,u_1,\tau)=-{2\pi \tau_1\over\vert\tau\vert^2}\left(u_1^2-u_1+{1\over 6}\right)
+\sum_{n=-\infty}^{\infty}\ln\left\vert 1-\exp\left[-2\pi i\left\{{\vert
n\vert\over \tau}+\epsilon_n\left({u_1\over\tau}-u_0\right)\right\}\right]
 \right\vert^2\no$$
Now to agree with the continuum calculation $F(u_0,u_1,\tau)$ should give the
$\bar\partial$-torsion  via the equation 
\eqlabel{\expressionone}
$$\ln T_{\bar\partial}(T^2,{\cal L})= F(u_0,u_1,\tau)\no$$
But $T_{\bar\partial}(T^2,{\cal L})$ was given in 
\docref{dbartorsion} from which  we obtain 
\eqlabel{\expressiontwo}
$$\eqalign{\ln T_{\bar\partial}(T^2,{\cal L})&=
\ln\left\vert\exp[\pi i u_0^2\tau]{\theta_1(u_1-\tau u_0,\tau)\over\eta(\tau)}\right\vert^2\cr
&=-2\pi \tau_1\left(u_0^2-u_0+{1\over 6}\right)
+\sum_{n=-\infty}^{\infty}\ln\left\vert 1-\exp\left[2\pi i\left\{\vert
n\vert \tau+\epsilon_n\left(u_1-u_0\tau\right)\right\}\right]
 \right\vert^2,\cr}\no$$
(The last line of \docref{expressiontwo} follows from the
 series representations of $\eta(\tau)$ and $\theta_1(u_1-\tau
u_0,\tau)$.)  However the apparent disagreement between
\docref{expressionone} and \docref{expressiontwo} is {\it illusory}
these two expressions are both {\it modular invariants} and if we
perform the summations leading to \docref{expressionone}  in the opposite order
then we get precisely \docref{expressiontwo}. Hence we do indeed find that the
continuum limit of our lattice model has reproduced the $\bar\partial$-torsion  
$T_{\bar\partial}(T^2,{\cal L})$ in a fashion quite analogous to the one dimensional 
example of the previous section.
\par 
It is useful at this point to
spell out some of the details of the modular properties of our
expressions in a short discussion.
We recall some basic facts. If $M$ denotes an element of the modular
group then $M$ acts on the modular parameter $\tau$
according to the well known rule $\tau\mapsto (a\tau+b)/(c\tau+d)$ with
integral $a$,$b$,$c$ and $d$ satisfying $ad-bc=1$. Hence we usually
write something like
$$M=\pmatrix{a&b\cr
c&d\cr}\in SL(2,{\bf Z})/{\bf Z}_2,\quad a,b,c,d\in{\bf Z},\; ad-cb=1\no$$
However when such transformations  $M$ act on the torus they act on
cycles and thus on the generators of the fundamental group $\pi_1(T^2)$;
or equivalently on  the holonomy phases $u_0$ 
and $u_1$ which specify the bundle ${\cal L}$. Hence we must work out
this action on the phases. But the modular group 
$SL(2,{\bf Z})/{\bf Z}_2 $ is generated by the celebrated $S$ and $T$
elements which are
$$S:\tau\longmapsto -{1\over\tau},\qquad T:\tau\longmapsto \tau+1\no$$
Therefore it is sufficient to give the action of $S$ and $T$ on the
holonomy phases and this is 
$$\eqalign{\hbox{Under } S&:u_0\longmapsto u_0,\qquad u_1\longmapsto
u_0+u_1\cr
\hbox{Under } T&:u_0\longmapsto u_1,\qquad u_1\longmapsto -u_0\cr}
          \no$$  
In addition $S$ and $T$ are isometries of the flat torus so they leave the volume
$V$ invariant.
\par
With this information on $S$ and $T$ it is possible to check directly the modular
invariance of \docref{expressionone} and \docref{expressiontwo}. 
For completeness we give the corresponding information for a general
element $M$ of the modular group. It is that, regarding 
$L_0$ and $L_1$ as complex numbers in an obvious way, we have 
$$\eqalign{\hbox{Under } M&:u_0\longmapsto c u_1+ d u_0,\qquad
u_1\longmapsto  a u_1+b u_0\cr
\hbox{and }&\tau\longmapsto {a \tau+b\over c\tau+d}\cr
\hbox{with } &M\in SL(2,{\bf Z})/{\bf Z}_2 \cr}\no$$ 
\par
For the $\bar\partial$-torsion  this modular invariance is the  assertion
that $T_{\bar\partial}(T^2,{\cal L})$ only depends on the conformal
structure on $T^2$ i.e. on its complex structure. In fact as observed in
\ref{10} modular invariance in this case can be shown to follow from the
classical functional equations obeyed by $\theta_1$ and $\eta$.
\par
In fact in this work the modular invariance
of the continuum theory is guaranteed by, and has its origin in, the
independence of the limit defining $F(u_0,u_1,\tau)$ on the order in which its  
double sums are carried out. 
\par 
Finally we reinstate the mass and give
the corresponding formulae when $m\not=0$. We find that
\eqlabel{\massnonzero} 
$$\eqalign{\ln\det{}_K&\left({\Delta+m^2\over2}\right) \rightarrow K^2
\int_{-\pi}^{\pi}{d\nu_1d\nu_2\over (2\pi)^2} \ln\left[\left\{
\delta-\alpha\cos(\nu_1)-\beta\cos(\nu_2)-\gamma\cos(\nu_1-\nu_2)
\right\}\right]\cr &-{\pi\tau_1\over6} c(u_0,{m^2V\over\tau_1})
+\sum_{n=-\infty}^{\infty} \ln\left\vert1-e^{-2\pi
\tau_1\sqrt{{(n+u_0)}^2+{m^2V\over4\pi^2\tau_1}}+2\pi
i\{u_1-\tau_0(n+u_0)\}}\right\vert^2 \cr}\no$$ where $V=L_0
L_1\sin\theta$ and the function $c(u,x)$ that appears in 
\docref{massnonzero} above is defined by the equation \eqlabel{\cdef}
$$\int_{-\infty}^{\infty}{dp\over2\pi}
\ln\left\vert1-e^{-\sqrt{p^2+x}+2\pi i u}\right\vert^2=
-{c(u,x)\pi\over6}\no$$ One can verify that \docref{massnonzero} 
reduces to \docref{expressiontwo} when $m=0$---for example this would
require that  $c(u,0)=12(u^2-u+1/6)$ which is true. 
\par 
Of course, for this case where  $m\not=0$ we can use a zeta function method to give 
the continuum partition function.
Repeated use of (\docref{planasum}) enables one to calculate 
$\zeta^\prime_{d^*_{\cal L}d_{\cal
L}+m^2}(0)$ which is the required
object. The result is that we obtain $W_F$ in agreement 
with \docref{massnonzero} above; we also find a bulk term of the 
form $-{Vm^2\over4\pi}(\ln[(m/2\pi\mu)^2]-1)$ where $\mu$ is 
an arbitrary undetermined scale.  
\par
We would like
to emphasize that, in the continuum or scaling
limit, we have found the exact finite size corrections to $\ln Z(T^2,{\cal L},m^2)$ 
and these are {\it modular invariant} whether $m=0$ or not. But when $m$
is zero  we have a critical phase and then the finite size corrections
to $Z$ exhibit  {\it holomorphic
factorisation }  and are given by elliptic theta functions, i.e.
$$ \eqalign{\exp[-W_F]&={\cal F}\longbar{\cal F}\cr
\hbox{with }{\cal F}&=\exp[\pi i u_0^2\tau]{\theta_1(u_1-\tau u_0,\tau)\over\eta(\tau)}\cr}\no$$
\par
Concerning holomorphic factorisation note that if the phase factor 
$\exp[\pi i u_0^2\tau]$ were absent from  ${\cal F}$ above then $Z$ would also exhibit  holomorphic factorisation in the Picard variety label 
$w=u_1-\tau u_0$. Actually the presence of the phase $\exp[\pi i u_0^2\tau]$ is a consequence of the central charge $c$ being non-zero---lack of holomorphic 
factorisation in this sense is the existence of a holomorphic anomaly 
\ref{12} of the $\bar\partial_{\cal L}$ determinant bundle over the Picard variety.
If we observe that 
$$\partial_{w}\partial_{\bar w}\ln T_{\bar\partial}(w,\bar w, \tau)
=4\pi\delta(w,\bar w)-{\pi\over\tau_1}\no$$
we see that the torsion is proportional to the Greens function, where the lowest mode (the zero mode) has been projected out, of the Laplacian $\partial_w\partial_{\bar w}$ on the 
Piccard variety. 
It is the absence of  this lowest mode that is the barrier to holomorphic factorization.
\par  
In the next
section we turn to the bulk term and the cylinder limit 
$L_1\rightarrow\infty$. 
\beginsection{Conformal properties and effects due to finite size and finite lattice
spacing.} In this section we work  with $K_0$, $K_1$ and the mass $m$ at
perfectly general values. 
\par 
We recall that $W$ and $Z$ are related by 
$$W=-\ln Z\no$$
If we now perform the lattice sums, doing first the sum over $n_0$,
followed by that over $n_1$ we obtain the result that
$$W=K_0K_1W_B+W_F\no$$ 
where  
$$W_B=\int_{-\pi}^{\pi}{d\nu_1d\nu_2\over
(2\pi)^2} \ln\left[{\sqrt{g}\over\pi}\left\{
\delta-\alpha\cos(\nu_1)-\beta\cos(\nu_2)-\gamma\cos(\nu_1-\nu_2)
\right\}\right]\no$$ 
and 
$$W_F=K_1\sum_{i=0}^1
\int_0^{\pi/2}{d\omega\over\pi}\, \ln\vert 1-(s^i_-)^{K_0}\vert^2
+\sum_{n_1=0}^{(K_1-1)}\ln\vert 1-v_{n_1}^{K_0}\vert^2\no$$ 
We note that
$W_B$ gives the free energy per lattice site in the thermodynamic limit,
i.e. we have 
$$\lim_{K_0,K_1\rightarrow\infty}{W \over K_0K_1}=W_B\no$$ 
This in turn means that $W_F$ gives the {\it complete finite size
corrections} to the bulk lattice behaviour. We now examine in more
generality and more detail the limits studied in the preceding section.
We now no longer set $K_0=K_1=K$ we simply take
$K_0,K_1\rightarrow\infty$ while keeping fixed their ratio $k=K_0/K_1$,
we also keep fixed the quantities $m^2$, $\theta$ and $L_i=K_0 a_i$, we
shall refer to this as the scaling limit. We still find the same result
for the finite size corrections $W_F$ namely 
$$\lim_{scaling}
W_F=-{\pi\tau_1\over6} c(u_0,{m^2V\over\tau_1})
+\sum_{n=-\infty}^{\infty} \ln\left\vert1-e^{-2\pi
\tau_1\sqrt{{(n+u_0)}^2+{m^2V\over4\pi^2\tau_1}}+2\pi
i\{u_1-\tau_0(n+u_0)\}}\right\vert^2 \no$$ while for the bulk term we
find the mass dependence $$\eqalign{\lim_{scaling}K_0 K_1 W_B&=K_0
K_1\Lambda_B-{Vm^2\over4\pi}\left\{\ln[K_0K_1]-2\rho\right\}
-{Vm^2\over4\pi}(\ln[{m^2V\over 4\pi^2}]-1)+\cdots\cr \hbox{where
}\Lambda_B&=W_B\vert_{m=0},\qquad V=L_0 L_1\sin\theta\cr \hbox{and
}\rho\equiv\rho(\alpha, \beta, \gamma)&=\int_{0}^{\pi}d\nu
\left[{1\over\sin\nu\sqrt{1+g\alpha^2\sin^2\nu}}-{1\over\nu}\right] 
-{1\over2}\ln\left[\sqrt{g}(\beta+\gamma)\right]\cr}\no$$ \par Our
results can be applied to the Ising model on a torus---the special case
corresponding, in our language, to $\tau_0=0$ was studied in \ref{13}.
This arises because of the equivalence of the Ising model and a dimer
model on a decorated lattice cf. \ref{14,15}. To see what happens we
denote $W_F$ by $W_F(u_0,u_1)$ in order to display its phase dependence
and use \ref{16} for $\alpha$, $\beta$, $\gamma$ and $\delta$. Let us do this:
denoting nearest neighbour spin-spin couplings by $J_1$, $J_2$ and $J_3$
then
\eqlabel{\Isingparametrization}
$$\eqalign{\alpha&=\sinh({2J_1\over k_B T}),\quad \beta=\sinh({2J_2\over k_B T}),
\quad \gamma=\sinh({2J_3\over k_B T})\cr
\delta&=\cosh({2J_1\over k_B T})\cosh({2J_2\over k_B T})\cosh({2J_3\over k_B T})
+\sinh({2J_1\over k_B T})\sinh({2J_2\over k_B T})\sinh({2J_3\over k_B
T})\cr}\no$$
and the Ising partition function is given by 
\eqlabel{\Zising}
$$Z^{Ising}={1\over2}e^{-W_{B}^{Ising}}\left\{ \mp
e^{{1\over2}W_F(0,0)}+e^{{1\over2}W_F(0,{1\over2})}+e^{{1\over2}W_F({1\over2},0)}
+e^{{1\over2}W_F({1\over2},{1\over2})}\right\}\no$$ This result is for
ferromagnetic couplings, with $+$ referring to $T<T_c$ and $-$ to
$T>T_c$.  If we take the  the scaling limit $W_F(u_0,u_1)\mapsto\Gamma_F(u_0,u_1)$
and we obtain a modular invariant expression for the finite size contributions in the 
scaling limit for this model also.  
Our results (\docref{Zising}) incorporates the complete
lattice and finite size corrections for the Ising model on a triangular
lattice. With a similar equivalence our results can easily be
translated to give the general  result for other models.  
\par
We point out that the expression of $Z^{Ising}$ as the four term sum
\docref{Zising} above constructs a modular invariant function of $\tau$
and $m^2V$; this construction makes it easy to understand that it is the
summing over the $u_i$ which form an orbit of  the $SL(2,{\bf Z})/{\bf Z}_2$-action  on the
space of $u_i$'s that guarantees the invariance. Mathematically we are
working with the action of $SL(2,{\bf Z})/{\bf Z}_2$ on the space of
flat bundles ${\cal L}$---the so called Picard variety---and many other
orbits exist, this allows us to construct many additional (phase
independent) modular invariant partition functions such as those of the
other conformally invariant field theories cf. \ref{17} and references therein.
Rational conformal theories will be obtained when the $u_i$ orbits
contain elements corresponding to roots of unity.
\par
Next we would like to consider an interesting geometric limit of the
model where $L_1\rightarrow\infty$: the cylinder limit. This will enable
us to access the central charge of the model. For large $L_1$ the
quantity $W_F/V$ tends to the finite value\footnote{$^{(d)}$}{\eightpoint
We could equally take $L_0$ large if we interchange $(u_0,L_0)$ with $(u_1,L_1)$.}
$\gamma^{cylinder}$  where 
\eqlabel{\cylin}
$$\gamma^{cylinder}=-{\pi\over 6L_0^2}c(u_0,m^2L_0^2)\no$$
We refer to $c(u_0,m^2L_0^2)$ as the `cylinder charge', the central charge $c$ of the model
is obtained by setting $m=u_0=0$ and
we see that this gives us
$$c=2\no$$
Because the cylinder charge determines the central charge at appropriate
values of its variables it might appear reasonable to expect that the cylinder charge
$\gamma^{cylinder}$ be equal to the Zamolodchikov $c$-function \ref{18}---this is
{\it not so.} To see this note that the $c$-function must be {\it
monotonic} in $m$ and the cylinder charge is not cf. fig. 2 where we
plot $c(u,x)$ versus $x$ for various $u$. In fig. 2 we see that $c(u,x)$
exhibits a universal crossover phenomenon as $x\rightarrow \infty$ with $c(u,x)$ having the
property that $\lim_{x\rightarrow\infty}c(u,x)\rightarrow 0$ whatever
the value of $u$. 
\par
The cylinder charge for the Ising model, can be obtained from
\docref{Zising} with $L_1\rightarrow\infty$, and it works out to be $\gamma^{cylinder}_{Ising}$ where 
$$\gamma^{cylinder}_{Ising}={\pi\over 12L_0^2}c({1\over
2},x)\no$$
This means that comparison with \docref{cylin} above determines that 
the cylinder charge for the Ising model is
$$-{1\over 2}c({1\over 2},x)\no$$
Now if we set $x=0$ we should get the Ising central charge and doing
this we find $c={1\over 2}$ (cf. fig. 2) the conventional value \ref{19,20}.
\par
A further rich feature of this model is that is possesses non-commuting
limits  namely the limits $u_0,u_1\rightarrow 0$ and $m\rightarrow 0$.
To establish this we can expand the finite size $W_F$ contribution to 
\docref{massnonzero} for small $u_i$ and $m$ obtaining
\eqlabel{\asymptotic}
$$W_F=\ln\left[(2\pi)^2\vert u_1-\tau u_0\vert^2+\tau_1 m^2V\right]
+2\ln\vert\eta(\tau)\vert^2+\cdots\no$$
But the RHS of \docref{asymptotic}  can tend to the two distinct
(logarithmically singular) expressions
$$\eqalign{&\ln\left\vert u_1-\tau
u_0\right\vert^2+2\ln\vert\eta(\tau)\vert^2\cr
 \hbox{and  } &\ln[\tau_1 m^2V]+2\ln\vert\eta(\tau)\vert^2\cr}\no$$ 
depending on the order in which the limits are taken. 
Nevertheless both limits and indeed 
\docref{asymptotic} itself are {\it modular invariant}. 
\par
This model is also ideal for studying the Kosterlitz-Thouless phase \ref{21}.
One can add on a $\vert\varphi\vert^4$ interaction and in this way one can 
study the critical point of the $XY$ model approaching it from the
disordered phase. We have holonomy conditions (\docref{bdryconds}) that describe the
essential features of a vortex phase: To make this more transparent one
should perform the cylinder limit of \docref{bdryconds} and then map the cylinder
to the plane using the conformal map
$$y+i{2\pi x\over L_0}=\ln z\no$$
This gives a vortex at the origin whose charge is given by $u_0$ and
whose presence  is detected by the existence of an Aharanov-Bohm effect
for transport around the origin.
\beginsection{Conclusion}
We have considered lattice models and their continuum limits in one and
two dimensions.  Our one dimensional  model is that of a  gauge
theory for a flat bundle $L$ over the circle and it is non-trivial its partition function being
a power of the Ray-Singer torsion $T(S^1,{\cal L})$ of ${\cal L}$. 
We find that the combinatorial and continuum partition functions agree precisely 
and are equal to $T(S^1,{\cal L})^{-1}$. This result emphasizes the topological nature of
$T(S^1,{\cal L})^{-1}$
which is of course expressible purely combinatorially, as it was
originally \ref{22}, or analytically as it was later \ref{23}.
\par
In summary, the finite size corrections to the free energy are {\it modular invariant.} 
This conclusion extends to the entire scaling neighborhood of the critical phase. 
We use our results to give expressions for the complete lattice and
finite size corrections for the two dimensional Ising model on a 
triangular lattice via its equivalence to a sum over Pfaffians. 
Modular invariance also extends to 
models in more than two dimensions when the geometry giving rise to finite size 
effects contains a flat torus. For a three dimensional 
cylindrical geometry with toroidal cross-section the result can be obtained from
$W_F$ by replacing $m^2$ with $m^2+q^2$ and integrating the resulting 
expression over $q$. One can understand the origin of modular invariance in general 
as the residual freedom to reparametrise coordinates, in the continuum limit,
while retaining flat toroidal geometry.
\par
In the two dimensional case the limiting finite size corrections 
at the critical phase are expressible 
in terms of classical elliptic functions. Infinitesimally small values of the phases 
$u_i$ lead to logarithmically  divergent contributions to the free energy. This 
means that the free energy needed to create a vortex 
becomes infinite for an infinitely large lattice.  In general the model has a 
surprisingly rich structure of non-commuting limits. For example the 
limits of approaching the critical phase ($m\rightarrow0$) and that of sending the $u_i$ to 
zero do not commute. 
\par\vfill\eject
\input epsf
\epsfxsize=0.6\hsize
\centerline{\bf The triangulated torus}
\par\vskip \baselineskip
\centerline{\epsffile{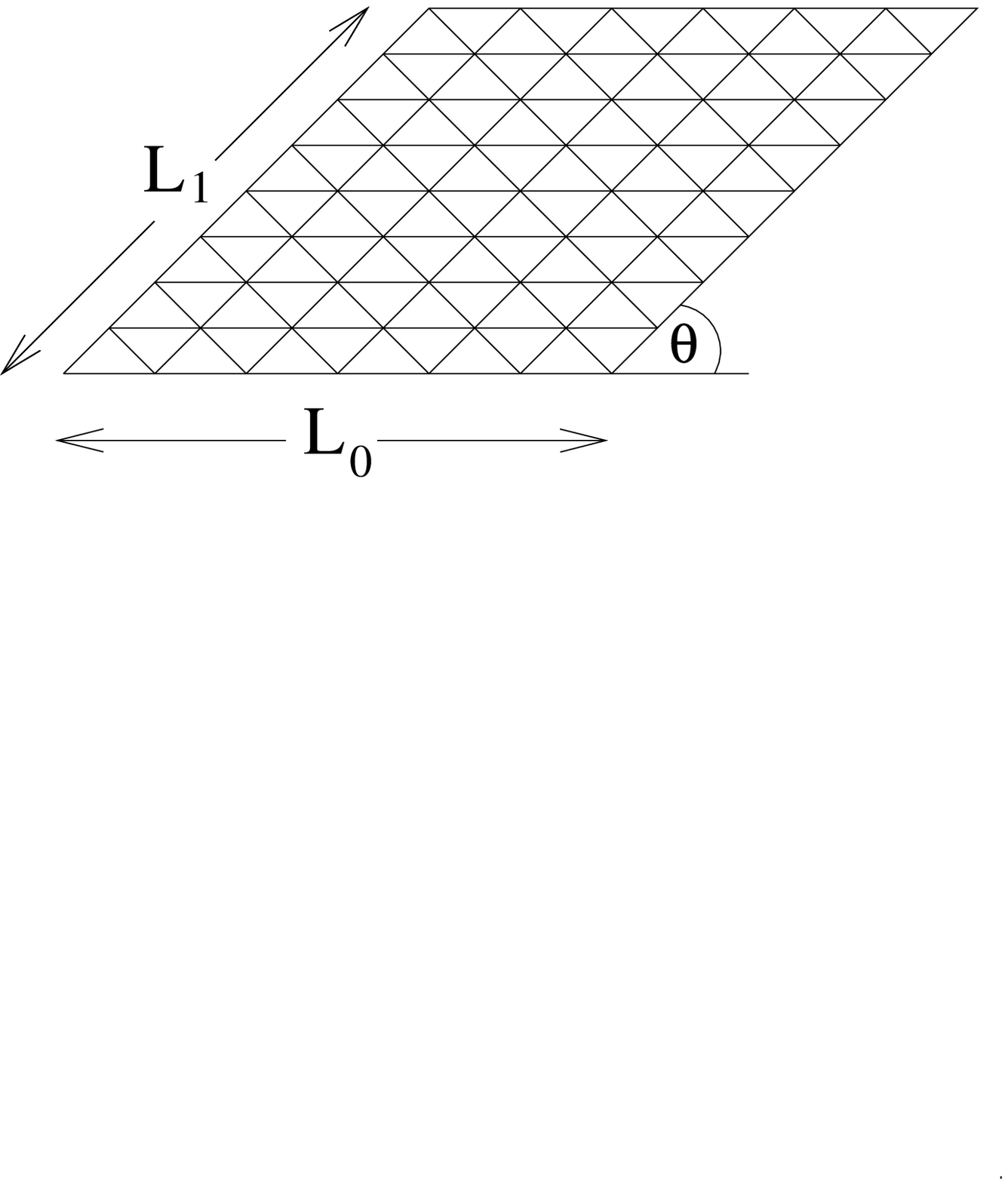}}
\par\vskip-0.25\vsize
\par\vskip-1.5\baselineskip
\centerline{\bf Fig. 1.}
\par\vskip \baselineskip
\centerline{\bf The `cylinder charge'  function $c(u,x)$ for various $u$}
\epsfxsize=0.6\hsize
\par\vskip-0.1\vsize
\centerline{\epsffile{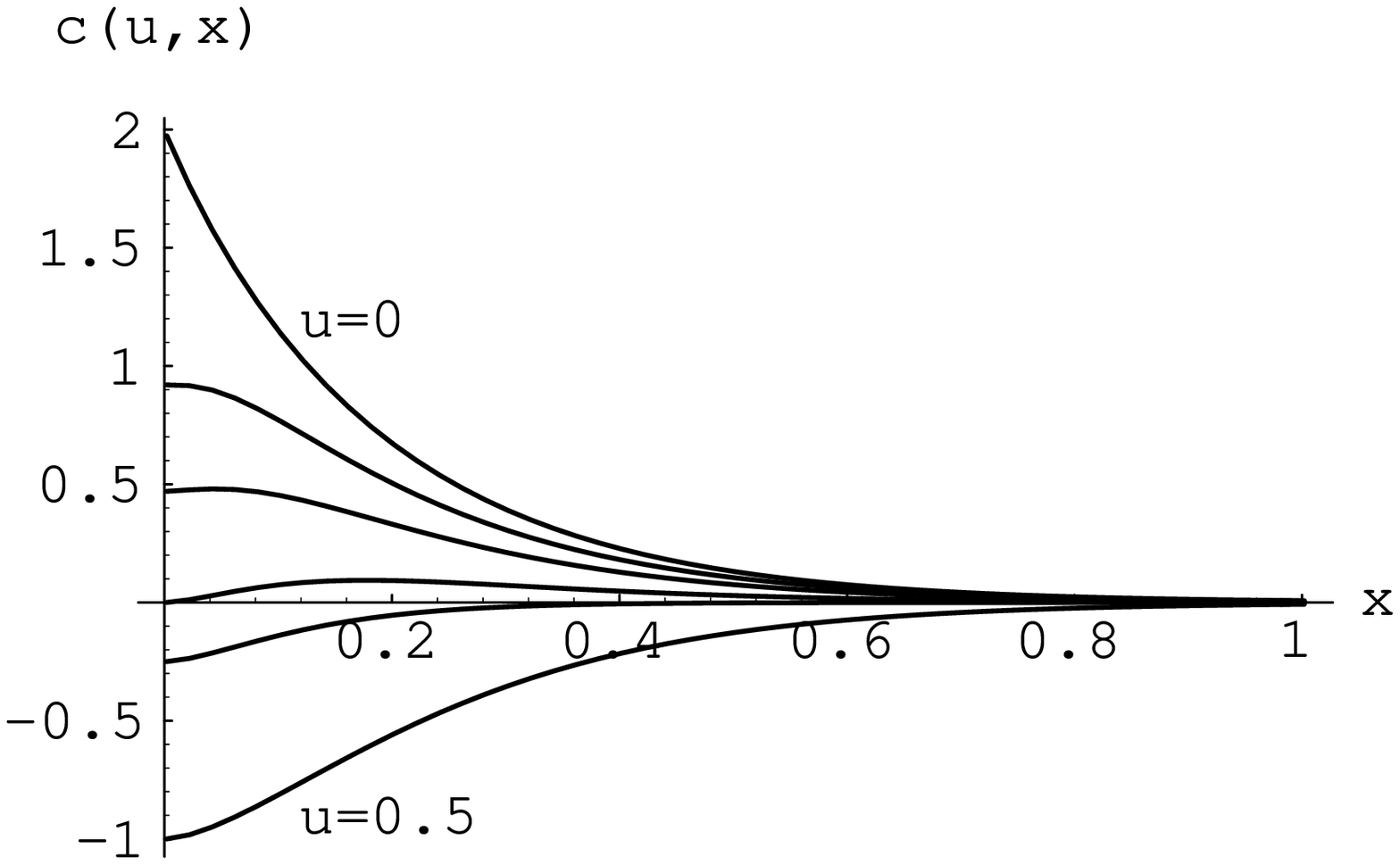}}
\par\vskip-1.5\baselineskip
\centerline{\bf Fig. 2.}
\par\vfill\eject
\centerline{\bf References}
\par
\vskip0.5\baselineskip
{ 
\par \noindent 
\par \hangindent 
\parindent \indent \hbox to\z@ {\hss \fam \bffam \tenbf 1.\kern .5em }\ignorespaces  Nash C. and O' Connor Denjoe, Modular invariance of finite size corrections and a vortex critical phase, Phys. Rev. Lett., {\fam \bffam \tenbf 76}, 1196--1199, (1996).
\par \vskip -0.8\baselineskip \noindent } 
{ 
\par \noindent 
\par \hangindent 
\parindent \indent \hbox to\z@ {\hss \fam \bffam \tenbf 2.\kern .5em }\ignorespaces  Barber M. N., {\fam \itfam \tenit Phase transitions and critical phenomena, vol. 8}, , {edited by: Domb C. and Lebowitz J. L.}, Academic Press, (1983). 
\par \vskip -0.8\baselineskip \noindent } 
{ 
\par \noindent 
\par \hangindent 
\parindent \indent \hbox to\z@ {\hss \fam \bffam \tenbf 3.\kern .5em }\ignorespaces  O' Connor Denjoe and Stephens C. R., Environmentally friendly renormalisation, Int. J. Mod. Phys. A, {\fam \bffam \tenbf 9}, 2805--2902, (1994).
\par \vskip -0.8\baselineskip \noindent } 
{ 
\par \noindent 
\par \hangindent 
\parindent \indent \hbox to\z@ {\hss \fam \bffam \tenbf 4.\kern .5em }\ignorespaces  Dotsenko V. S. and Fateev V. A., Conformal algebra and multipoint correlation functions in 2D statistical models, Nucl. Phys., {\fam \bffam \tenbf B240}, 312--348, (1984).
\par \vskip -0.8\baselineskip \noindent } 
{ 
\par \noindent 
\par \hangindent 
\parindent \indent \hbox to\z@ {\hss \fam \bffam \tenbf 5.\kern .5em }\ignorespaces  Dotsenko V. S. and Fateev V. A., Four point correlation functions and the operator algebra in 2D conformal invariant theories with central charge $c\le 1$, Nucl. Phys., {\fam \bffam \tenbf B251}, 691--734, (1985).
\par \vskip -0.8\baselineskip \noindent } 
{ 
\par \noindent 
\par \hangindent 
\parindent \indent \hbox to\z@ {\hss \fam \bffam \tenbf 6.\kern .5em }\ignorespaces  Nash C. and O' Connor Denjoe, BRST quantisation and the product formula for the Ray-Singer torsion, Int. Jour. Mod. Phys. A, {\fam \bffam \tenbf 10}, 1779--1805, (1995).
\par \vskip -0.8\baselineskip \noindent } 
{ 
\par \noindent 
\par \hangindent 
\parindent \indent \hbox to\z@ {\hss \fam \bffam \tenbf 7.\kern .5em }\ignorespaces  Nash C. and O' Connor Denjoe, Determinants of Laplacians, the Ray-Singer Torsion on Lens Spaces and the Riemann zeta function, Jour. Math. Physics, {\fam \bffam \tenbf 36}, 1462--1505, (1995).
\par \vskip -0.8\baselineskip \noindent } 
{ 
\par \noindent 
\par \hangindent 
\parindent \indent \hbox to\z@ {\hss \fam \bffam \tenbf 8.\kern .5em }\ignorespaces  Whittaker E. T. and Watson G. N., {\fam \itfam \tenit A course of modern analysis}, Cambridge University Press, (1927). 
\par \vskip -0.8\baselineskip \noindent } 
{ 
\par \noindent 
\par \hangindent 
\parindent \indent \hbox to\z@ {\hss \fam \bffam \tenbf 9.\kern .5em }\ignorespaces  Cheeger J., Analytic torsion and the heat equation, Ann. Math., {\fam \bffam \tenbf 109}, 259--322, (1979).
\par \vskip -0.8\baselineskip \noindent } 
{ 
\par \noindent 
\par \hangindent 
\parindent \indent \hbox to\z@ {\hss \fam \bffam \tenbf 10.\kern .5em }\ignorespaces  Ray D. B. and Singer I. M., Analytic Torsion for complex manifolds, Ann. Math., {\fam \bffam \tenbf 98}, 154--177, (1973).
\par \vskip -0.8\baselineskip \noindent } 
{ 
\par \noindent 
\par \hangindent 
\parindent \indent \hbox to\z@ {\hss \fam \bffam \tenbf 11.\kern .5em }\ignorespaces  Patrick A. E., On phase separation in the spherical model of a ferromagnet: quasiaverage approach, Jour. Stat. Phys., {\fam \bffam \tenbf 72}, 665--701, (1993).
\par \vskip -0.8\baselineskip \noindent } 
{ 
\par \noindent 
\par \hangindent 
\parindent \indent \hbox to\z@ {\hss \fam \bffam \tenbf 12.\kern .5em }\ignorespaces  Nash C., {\fam \itfam \tenit Differential Topology and Quantum Field Theory}, Academic Press, (1991). 
\par \vskip -0.8\baselineskip \noindent } 
{ 
\par \noindent 
\par \hangindent 
\parindent \indent \hbox to\z@ {\hss \fam \bffam \tenbf 13.\kern .5em }\ignorespaces  Ferdinand A. E. and Fisher M. E., Bounded and inhomogeneous Ising models I. Specific heat anomaly of a finite lattice, Phys. Rev., {\fam \bffam \tenbf 185}, 832--846, (1969).
\par \vskip -0.8\baselineskip \noindent } 
{ 
\par \noindent 
\par \hangindent 
\parindent \indent \hbox to\z@ {\hss \fam \bffam \tenbf 14.\kern .5em }\ignorespaces  Kastelyn P. W., Dimer statistics and phase transitions, J. Math. Phys., {\fam \bffam \tenbf 4}, 287--293, (1963).
\par \vskip -0.8\baselineskip \noindent } 
{ 
\par \noindent 
\par \hangindent 
\parindent \indent \hbox to\z@ {\hss \fam \bffam \tenbf 15.\kern .5em }\ignorespaces  Fisher M. E., On the dimer solution of planar Ising models, J. Math. Phys., {\fam \bffam \tenbf 7}, 1776--1781, (1966).
\par \vskip -0.8\baselineskip \noindent } 
{ 
\par \noindent 
\par \hangindent 
\parindent \indent \hbox to\z@ {\hss \fam \bffam \tenbf 16.\kern .5em }\ignorespaces  Stephenson J., Ising-Model spin correlations on the triangular lattice, J. Math. Phys., {\fam \bffam \tenbf 5}, 1009--1024, (1964).
\par \vskip -0.8\baselineskip \noindent } 
{ 
\par \noindent 
\par \hangindent 
\parindent \indent \hbox to\z@ {\hss \fam \bffam \tenbf 17.\kern .5em }\ignorespaces  Itzykson C. and Drouffe J., {\fam \itfam \tenit Statistical field theory, vol. II}, Cambridge University Press, (1989). 
\par \vskip -0.8\baselineskip \noindent } 
{ 
\par \noindent 
\par \hangindent 
\parindent \indent \hbox to\z@ {\hss \fam \bffam \tenbf 18.\kern .5em }\ignorespaces  Zamolodchikov A. B., Irreversibility of the flux of the renormalisation group in a 2d field theory, JETP Lett., {\fam \bffam \tenbf 43}, 730--732, (1986).
\par \vskip -0.8\baselineskip \noindent  } 
{ 
\par \noindent 
\par \hangindent 
\parindent \indent \hbox to\z@ {\hss \fam \bffam \tenbf 19.\kern .5em }\ignorespaces  Bl{\accent "7F o}te H. W. J., Cardy J. L. and Nightingale M. P., Conformal invariance, the central charge and universal finite-size amplitudes at criticality, Phys. Rev. Lett., {\fam \bffam \tenbf 56}, 742--745, (1986).
\par \vskip -0.8\baselineskip \noindent } 
{ 
\par \noindent 
\par \hangindent 
\parindent \indent \hbox to\z@ {\hss \fam \bffam \tenbf 20.\kern .5em }\ignorespaces  Affleck I., Universal term in the free energy at a critical point and the conformal anomaly, Phys. Rev. Lett., {\fam \bffam \tenbf 56}, 746--748, (1986).
\par \vskip -0.8\baselineskip \noindent } 
{ 
\par \noindent 
\par \hangindent 
\parindent \indent \hbox to\z@ {\hss \fam \bffam \tenbf 21.\kern .5em }\ignorespaces  Kosterlitz J. M. and Thouless D. J., Ordering, metastability and phase transitions in two dimensional systems, J. Phys. C, {\fam \bffam \tenbf 6}, 1181--1203, (1973).
\par \vskip -0.8\baselineskip \noindent } 
{ 
\par \noindent 
\par \hangindent 
\parindent \indent \hbox to\z@ {\hss \fam \bffam \tenbf 22.\kern .5em }\ignorespaces  Franz W., {\accent "7F U}ber die Torsion einer {\accent "7F U}berdeckung, J. Reine Angew. Math., {\fam \bffam \tenbf 173}, 245--254, (1935).
\par \vskip -0.8\baselineskip \noindent } 
{ 
\par \noindent 
\par \hangindent 
\parindent \indent \hbox to\z@ {\hss \fam \bffam \tenbf 23.\kern .5em }\ignorespaces  Ray D. B. and Singer I. M., R-torsion and the Laplacian on Riemannian manifolds, Adv. in Math., {\fam \bffam \tenbf 7}, 145--201, (1971).
\par \vskip -0.8\baselineskip \noindent } 
\bye
\cite{781} 
\cite{789} 
\cite{784} 
\cite{785} 
\cite{764} 
\cite{673} 
\cite{664} 
\cite{184} 
\cite{765} 
\cite{780} 
\cite{786} 
\cite{787} 
\cite{788} 
\cite{783} 
\cite{7}   
\cite{286} 
\cite{779} 
\cite{777} 
\cite{670} 
\cite{183} 
\cite{626} 
\bye